\documentclass [a4paper,12pt]{article}
\usepackage{amsmath,amssymb,graphicx,cite,delarray}
\usepackage[a4paper]{geometry}
\usepackage[bf]{caption2}

\begin{document}
\pagestyle{plain}
\sloppy
\begin{center}\begin{Large}THE MATERIAL ORIENTATION RELATIONSHIP FOR THE BCC-HCP TRANSITION\end{Large}\end{center}

\begin{center}
M.P. Kashchenko, V.G.Chashchina\end{center}

\begin{center}
Physics Chair,  Ural State Forest Engineering University,
Sybirskiy trakt, 37, 620100, Ekaterinburg, Russia
mpk46@mail.ru\end{center}

\begin{abstract} The dynamical model  of  forming of  martensitic crystals  for the bcc-hcp  transition is offered. It is shown that all macroscopic morphological characters (the habit plane, the macroshear and the orientational relationship) are expressed through elastic moduluses Cij of an initial bcc phase. 
\end{abstract}

\section{Introduction}
 
Earlier for fcc-bcc martensitic transformation the model of heterogeneous nucleation (in elastic fields of dislocations) and wave growth of the martensite was formulated. 

The elastic field of dislocations disturbings the original lattice symmetry by selecting regions being most favorable for martensitic nucleation. 

Such a region features the shape of a perpendicular parallelepiped, its edges being oriented along the eigenvectors  $\mathbf{\xi}_{i}$ of the strain tensor $\hat{\varepsilon}$, its eigenvalues $\varepsilon_{i}$ satisfying the conditions:
\begin{equation}
\varepsilon_{1}>0, \quad \varepsilon_{2}<0, \quad \varepsilon_{3} \approx 0.
\label{eq1}
\end{equation}

The transformation starts with the emergence of an excited state with the shape of a parallelepiped, its pairs of faces are oscillating in opposed phase, thereby  exciting controlling displacement waves orientate in the wave-normal $\mathbf{n}_{1,2}$ close to $\mathbf{\xi}_{1,2}$. In the most simple approximation: 
\begin{equation}
\mathbf{n}_{1} =  \mathbf{\xi}_{1},\quad \mathbf{n}_{2} =  \mathbf{\xi}_{2},\quad  \vert \mathbf{n}_{1,2} \vert = \vert \mathbf{\xi_{1,2}} \vert = 1.	
\label{eq2}                                                                      
\end{equation}
                                                                                                                  
\begin{figure}[htb] 
\centering
\includegraphics[clip=true,width=.8\textwidth]{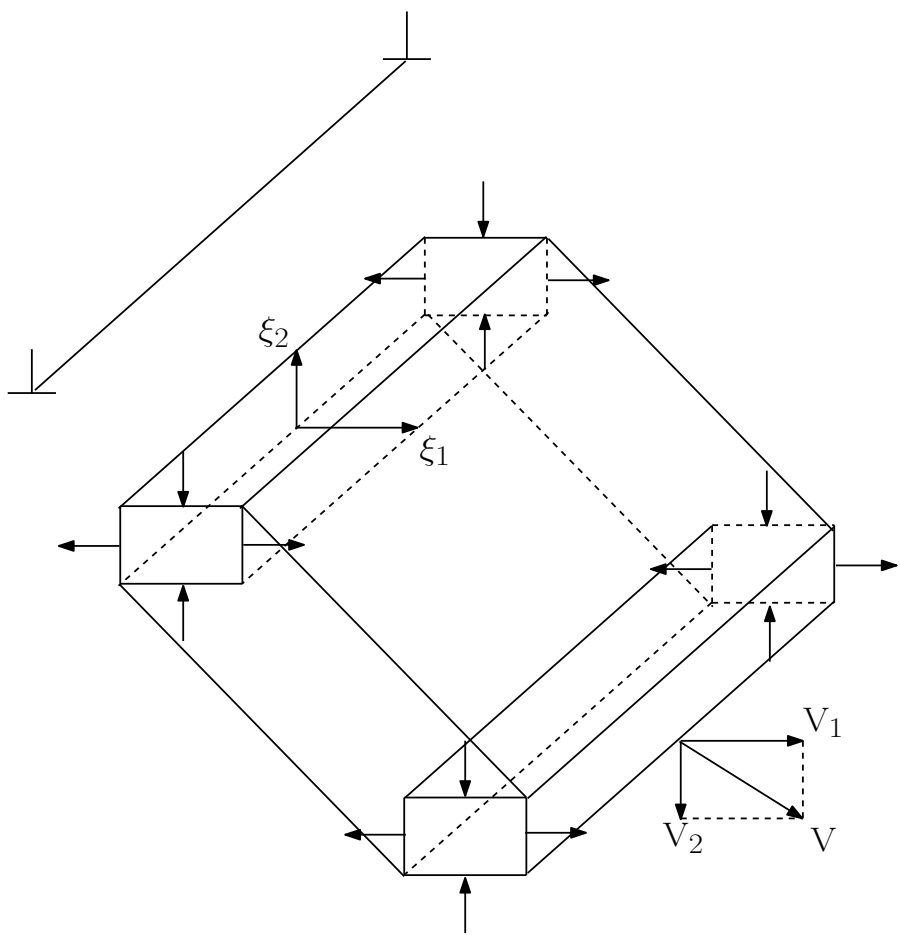}
\caption{The controlling wave process by the growth of the martensitic lamella}
\label{fig1}
\end{figure}

According to the wave model \cite{bib:1} of the martensite crystal growth (Fig.\ref{fig1}) in the interphase region there is the threshold flat strain of the tension ($\varepsilon_{1th}>0$)-compression ($\varepsilon_{2th}<0$) type. On the mesoscale within the  crystal thickness  this threshold strain is uniform.

This model gives natural interpretation of the supersonic velocity \textbf{v} of a separate crystal growth. Let's note that the given variant sets an upper limit of growth speed of a crystal for quite definite dynamic structure of the interphase region. For other types of martensitic transformations (at other dynamic structure of the wave front) growth speed can be less. The exceeding of threshold strain should lead to the loss of stability of an initial phase. Components of finite strain disturbing the symmetry of an initial lattice can achieve values $\sim 0.1$ that are typical for the Bein's strain. As shown earlier in the interphase region strongly non-equilibrium conditions are realized.
 
Controlling wave process (CWP) sets the lamellar form of crystals and the normal vector $\mathbf{N}$ to a habit plane of a growing crystal:
\begin{equation}
\mathbf{N}\, \Vert\, \mathbf{N}_{w}\,\Vert\, \mathbf{n}_{2}-\mathbf{n}_{1}\varkappa.                                                                                                         
\label{eq3}
\end{equation}

It is important that connection between absolute values of deformations and speeds $\mathrm{v}_{1}$ and $\mathrm{v}_{2}$ of elastic waves takes place
\begin{equation}
\varkappa = \Big{\vert} \frac{\mathrm{v}_{2}}{\mathrm{v}_{1}} \Big{\vert} = k = \frac{1-\vert \varepsilon_{2} \vert}{1+\varepsilon_{1}} \Big{(}\frac{\varepsilon_{1}(2+\varepsilon_{1})}{\vert \varepsilon_{2} \vert (2-\vert \varepsilon_{2} \vert)} \Big{)}^{\frac{1}{2}}.
\label{eq4}   
\end{equation}

At the threshold deformation $\varepsilon_{1},\vert \varepsilon_{2} \vert \ll 1$  therefore in \eqref{eq4} it is possible to believe  
\begin{equation}
k_{th}  \approx \Big{(}\frac{\varepsilon_{1}}{\vert \varepsilon_{2} \vert}\Big{)}_{th} \approx  \varkappa^{2}  =  \Big{\vert} \frac{ \mathrm{v}_{2}}{\mathrm{v}_{1}} \Big{\vert}^{2}.  
\label{eq5}                                                              
 \end{equation}      
         
\section{The role of CWP during bcc-hcp transformation}

In the case of the bcc-hcp  transformation the preference are natural for giving CWP stimulating the fastest transformation of $\{ 110 \}_{b}$ - planes of a cubic lattice to basic $\{ 0001 \}_{h}$ - planes of hcp lattice \cite{bib:2}. It is achieved if the wave process carries the deformation of compression along a direction $\langle 001 \rangle _{b}$ (and the deformation of tension along a direction $\langle 110 \rangle _{b}$):
 \begin{equation}
\mathbf{n}_{1} =  \mathbf{\xi}_{1} \, \Vert \,\langle 110 \rangle _{b}, \quad      \mathbf{n}_{2} =  \mathbf{\xi}_{2}  \,\Vert\, \langle 001 \rangle _{b}. 
\label{eq6}
\end{equation}       

After the loss of stability of the bcc - lattice the uniform (on the scale of the crystal thickness) deformation changes a nuclear potential relief, inducing the instability to the shear of either of the second plane.  This shear finishes the formation of the hcp - lattice.  
In our opinion, there is one more key idea. Namely:  the lamellar area of a lattice (which have lost the stability), aspiring in the constrained conditions to new symmetry of an arrangement of atoms, keeps the constant ratio of the main deformations given by the CWP.  How it follows from \eqref{eq5}, this ratio is equal $\varkappa^{2}$. Then, having the directions $\mathbf{\xi}_{i}$ calculated preliminary, it is easy to express through the same parameter $\varkappa$ all morphological attributes (the habit plane, the macroshear, the interphase orientational relationship, the relative change of volume and even the parameter of a hcp-lattice). The detailed analysis  of the bcc-hcp  transformation on an example of the titanium confirms this conclusion.

Alongside with \eqref{eq5} and \eqref{eq6} the condition of transformation of a plane $(110)_{b}$ in a plane $(0001)_{h}$ is essential also:      
\begin{equation}
(1-\vert \varepsilon_{2}\vert) \sqrt{3} = (1+\varepsilon_{1}) \sqrt{2}.
\label{eq7}
\end{equation}

 The values (in GPa) of elastic modules for bcc - Ti were used at quantitative estimations: $C_{11}=134, \,C_{12}=110, \,C_{44}=36$. For   $\mathbf{n}_{1}\, \Vert \, [110]_{b}$  and  $\mathbf{n}_{2}\, \Vert \, [001]_{b}$ from \eqref{eq5}  it is obtained:
\begin{equation}
\Big{(}\frac{\varepsilon_{1}}{\vert\varepsilon_{2}\vert}\Big{)}_{th} \simeq \varkappa^2 =\frac{2C_{11}}{C_{11}+C_{12}+2C_{44}}.
\label{eq8}
\end{equation}

In particular from \eqref{eq8} for bcc - Ti it is obtained that  $\Big{(}\frac{\varepsilon_{1}}{\vert\varepsilon_{2}\vert}\Big{)}_{th} \simeq 0.8481$. 

\section{The interphase orientational relationship} 

As is known there exists the interphase orientational relationship (OR). Usually  OR indicate  parallelism (or approximate parallelism) among the densely packed phase-planes and among angles of rotation in relation to the orientation of the densely packed directions possessed by  the parallel planes. In particular for the bcc- hcp martensitic transformation it is natural to believe that the plane which has tested the fastest transformation is included in OR:
\begin{equation}
\{110\}_{c} \, \Vert\, \{0001\}_{h}. 
\label{eq9}
\end{equation}

\begin{figure}[htb] 
\centering
\includegraphics[clip=true,width=.8\textwidth]{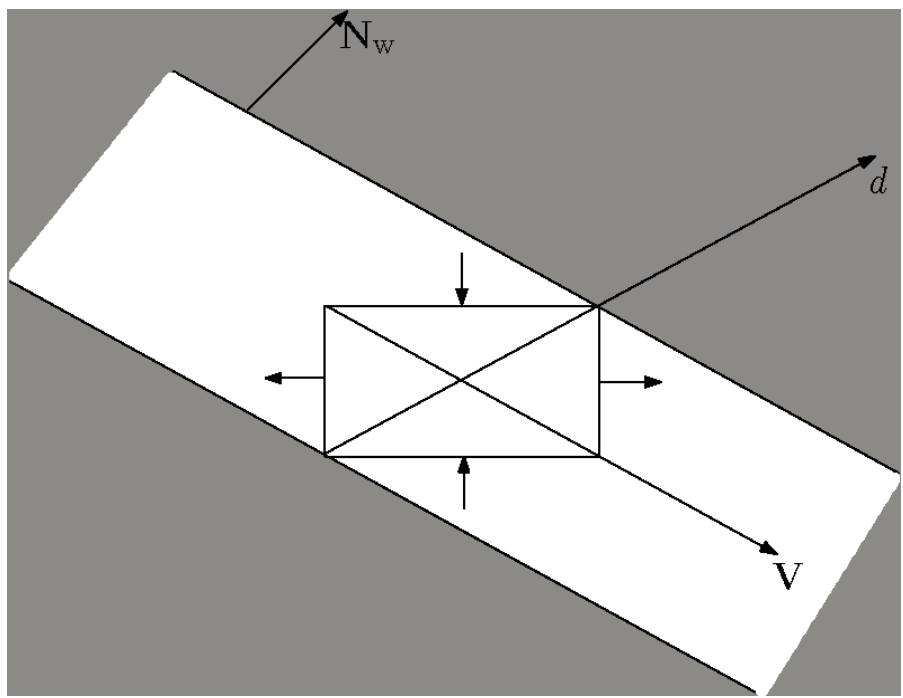}
\caption{The formation of the martensitic reaction channel  by propagating threshold deformations.}
\label{fig2}
\end{figure}

It is necessary to find only the conformity between the densely packed directions laying in planes. For this purpose it is enough to calculate the angle of the turn  of the orthogonal axes $\langle \bar{2} 110\rangle_{h}$, $\langle 0 1\bar{1} 0\rangle_{h}$ to the orthogonal axes $\langle 001\rangle_{c}$, $\langle 110\rangle_{c}$. It is necessary to take into account, that the turning material actually appears in the flat "channel" formed by the CWP in the bulk of the metastable austenite. Then the formation of a new phase with inevitability results in the turn of a lattice. The blacked out area on fig.~\ref{fig2} corresponds to the metastable austenite. The flat of the fig.~\ref{fig2} is coincident with the plane $(1\bar{1}0)_{c}$. Axes of the main deformations correspond to vertical and horizontal directions. 
 
The rectangular with arrows corresponds to (at some moment of time $t_{0}$) the position of section of the rectangular parallelepiped forming (during propagating) the area (light area on fig.~\ref{fig2}) playing a role of the flat "channel" inside of the  austenite. From fig.~\ref{fig2} it is obvious, that the section of "channel" is consistently "swept up" by the diagonal (along the vector d) of the rectangular  concerning both walls of "channel". 

\begin{figure}[!b] 
\centering
\includegraphics[clip=true,width=.7\textwidth]{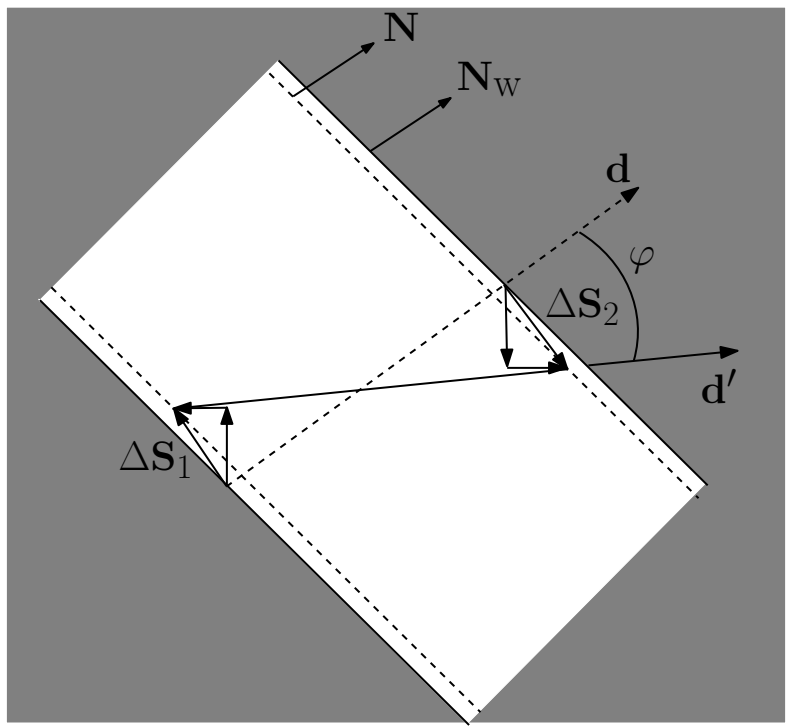}
\caption{The angle of the turn of a transforming lattice at final deformations.}
\label{fig3}
\end{figure}

On fig.~\ref{fig3} the increased fragment of a  transforming material containing the diagonal is represented. The diagonal at the moment $t_{0}$ (along a vector $\mathbf{d}$) corresponds  to threshold deformations $(\sim 10^{-3}\div10^{-4})$.  The diagonal at the moment $t_{0}+\Delta t$ (along a vector $\mathbf{d}^{\prime}$) corresponds to the end of the compression - tension process for  a  transforming material. The vectors $\Delta \mathbf{S}_{1,2}$ correspond  to displacements of boundary points of a diagonal.  They are shown as the sums of final displacements along the main axes of deformation. It is obvious that final position of a diagonal (with the changed length) corresponds to the turn in the constrained conditions on an angle $\varphi$ around of an axis $[1\bar{1}0]_{c}$.  The dashed lines, being parallel to initial boundaries of "channel", show final positions of boundaries (under condition of $\vert \varepsilon_{2}\vert < \varepsilon_{1}$).

The value of an angle   is calculated from condition
\begin{equation}
\cos{\varphi} = \frac{(\mathbf{d},\mathbf{d}^{\prime})}{d\, d^{\prime}},
\label{eq10}
\end{equation}
where (\quad ,\quad) is a symbol of the scalar product of vectors and

\begin{equation} 
\mathbf{d}\,\Vert\, \mathbf{v}_{1}-\mathbf{v}_{2}  = \mathbf{n}_{1} \cdot \mathrm{v}_{1} -  \mathbf{n}_{2} \cdot \mathrm{v}_{2},
\label{eq11}
\end{equation}

\begin{equation} 
\mathbf{d}^{\prime} \,\Vert \,\mathbf{v}_{1}(1 +  \varepsilon_{1}) - \mathbf{v}_{2} (1-\vert \varepsilon_{1}\vert) =  \mathbf{n}_{1} \mathrm{v}_{1}(1 +  \varepsilon_{1}) -  \mathbf{n}_{2} \mathrm{v}_{2} (1-  \vert \varepsilon_{1}\vert). 
\label{eq12}
\end{equation} 

Substituting \eqref{eq12} and \eqref{eq11} in \eqref{eq10} and taking into account \eqref{eq7} for ideal bcc-hcp martensitic transformation it is finally obtained:
\begin{equation}
 \cos{\varphi(\varkappa)} = \frac{\sqrt{\frac{3}{2}}+\varkappa^{2}}{\sqrt{\Big{(} \frac{3}{2}+\varkappa^{2}\Big{)}\Big{(} \varkappa^{2}+1\Big{)}}}.
\label{eq13}
\end{equation}

Thus, from the dynamic model of the martensite crystal growth  it is following  the general OR: 
\begin{displaymath}
\{110\}_{c}\,\Vert\, \{0001\}_{h},
\end{displaymath}
 
\begin{equation}
\langle 001\rangle _{c}\wedge  \langle \bar{2}110 \rangle _{h} =  \langle 110\rangle _{c}\wedge  \langle 01\bar{1}0 \rangle _{h}  = \varphi(\varkappa) = \arccos{\frac{\sqrt{\frac{3}{2}}+\varkappa^{2}}{\sqrt{\Big{(} \frac{3}{2}+\varkappa^{2}\Big{)}\Big{(} \varkappa^{2}+1\Big{)}}}}. 
\label{eq14}
\end{equation}

As far as it is known to authors, the analytical dependence of OR on the elastic properties of the material (reflected by the parameter $\varkappa$ \eqref{eq5}) is obtained for the first time.

Using the relationship \eqref{eq14} it is possible to present OR in other form:
\begin{equation}
\{110\}_{c}  \,\Vert \,\{0001\}_{h}, \quad [11\bar{1}]_{c}\wedge [11 \bar{2}0]_{h} = \varphi (\varkappa)- \varphi_{0},\quad \varphi_{0} = \arccos{\frac{1+\sqrt{6}}{2\sqrt{3}}}\approx 5. 26439^{\circ},
\label{eq15}
\end{equation}
where $[11\bar{1}]_{c}\wedge [11 \bar{2}0]_{h}$ is the angle between the nearest densely packed directions possessed by  the parallel planes (\ref{eq9}).

At $\varphi (\varkappa) - \varphi_{0} = 0$ the orientational relationship (\ref{eq15}) are named  Burgers OR. According to (\ref{eq14}) and (\ref{eq15}), Burgers OR is carried out only at $\varkappa^{-2} = 2$. Using the data for the titanium, we find: $\varphi(\varkappa) - \varphi_{0} \approx  0.43827^{\circ}$. Taking into account, that OR \eqref{eq14} and \eqref{eq15} obviously include elastic properties of a material, in our opinion, these OR pertinently to name as "material orientational relationship".
  
\section{Conclusion}

For the first time in the theory of  reconstructive martensitic transformations on the basis of the dynamic approach  the algorithm of the description of OR is established. OR are submitted as analytical dependences on elastic properties of an initial phase. The similar conclusions can be made for all morphological characteristics.   Hence the controlling wave process (stimulating the fastest way of martensitic  transformation) predetermines final morfological characteristics. Thus the depth of the   controlling covers a wide interval of deformations $(\sim 10^{-4}\div 10^{-1})$.

\end{document}